\pdfoutput=1
\documentclass[reprint,superscriptaddress,amsmath,amssymb,aps,pre]{revtex4-1}
\usepackage[caption=false]{subfig}
\usepackage{float}
\usepackage{amsmath}
\usepackage{natbib}
\usepackage{graphicx}
\usepackage{dcolumn}
\usepackage{bm}

\begin{document}

\title{Scaling law of the drag force in dense granular media}
\author{Sonu Kumar}
\affiliation{Department of Chemical Engineering, Indian Institute of Technology, Guwahati, Assam, India}
\author{K. Anki Reddy}
\email{anki.reddy@iitg.ernet.in}
\affiliation{Department of Chemical Engineering, Indian Institute of Technology, Guwahati, Assam, India}
\author{Satoshi Takada}
\affiliation{Earthquake Research Institute, University of Tokyo, 1-1-1, Yayoi, Bunkyo-ku, Tokyo 113-0032, Japan}
\affiliation{Yukawa Institute for Theoretical Physics, Kyoto University, Kitashirakawa Oiwakecho, Sakyo-ku, Kyoto 606-8502, Japan}
\author{Hisao Hayakawa}
\email{hisao@yukawa.kyoto-u.ac.jp}
\affiliation{Yukawa Institute for Theoretical Physics, Kyoto University, Kitashirakawa Oiwakecho, Sakyo-ku, Kyoto 606-8502, Japan}

\date{\today}

\begin{abstract}
Making use of the system of pulling a spherical intruder in static three-dimensional granular media, we numerically study the scaling law for the drag force $F_{\rm drag}$ acting on the moving intruder under the influence of the gravity.
Suppose if the intruder of diameter $D$ immersed in a granular medium consisting of grains of average diameter $d$ is located at a depth $h$ and moves with a speed $V$, we find that $F_{\rm drag}$ can be scaled as $(D+d)^{\phi_\mu} h^{\alpha_\mu}$ with two exponents $\phi_\mu$ and $\alpha_\mu$, which depend on the friction coefficient $\mu$ and satisfy an approximate sum rule $\phi_\mu+\alpha_\mu\approx 3$.
This scaling law is valid for the arbitrary Froude number (defined by $\mathrm{Fr}={2 V}\sqrt{{2D}/{g}}\big/(D+d)$), if $h$ is sufficiently deep.
We also identify the existence of three regimes (quasistatic, linear, and quadratic) at least for frictional grains in the velocity dependence of drag force.
The crossovers take place at $\mathrm{Fr}\approx 1$ between the quasistatic to the linear regimes and at $\mathrm{Fr}\approx 5$ between the linear to the quadratic regimes.
We also observe that Froude numbers at which these crossovers between the regimes happen are independent of the depth $h$ and the diameter of the intruder $D$.
We also report the numerical results on the average coordination number of the intruder and average contact angle as functions of intruder velocity.
\end{abstract}

\maketitle

\section{Introduction}\label{sec:intro}
Granular medium, an assembly of discrete particles behaves as a solid or a fluid depending on its density \cite{jaeger96}.
Variety of interesting phenomena exhibited by granular materials \cite{aranson06} has attracted many physicists and engineers in the last few decades.
There are many studies on the motion of the objects in granular media (through analysis of drag and lift forces) to improve the understanding of rheology of granular flows.

There are both experimental and simulation studies related to the characterization of the drag force on a moving passive object.
In most of these studies the objects considered are either spheres \cite{hilton}, cylinders \cite{reddy11,albert99,chehata03,wassgren03, bharadwaj06,guillard13,guillard15, zhang15}, or plate-like objects \cite{gravish10,guo12}.
Through the analysis of slow drag on a cylinder in aluminium oxide polydisperse granular mixtures under high pressure \cite{zhou05}, it is clarified that the drag resistance in granular media is related to the packing effects.
The studies of drag in a quasi-two-dimensional dilute supersonic granular flow \cite{boudet10} and in monolayered (two-dimensional) dense granular media \cite{takehara10} have provided more quantitative descriptions of drag force as the effects of flow fields around the intruder.
A quite recent study on the drag force of a spherical intruder moving through sedimented granular hydrogels \cite{panaitescu17} reports that drag force is a constant up to a critical velocity and starts increasing with velocity quadratically.
In a study of granular drag inspired by self-burrowing rotary seeds \cite{jung17}, the drag reduction by rotation for a vertically penetrating intruder is observed.
Similarly the drag reduction by mechanical vibration for the penetration of an intruder into a dry granular medium has been reported \cite{texier17} and this is attributed to the local fluidization which could rupture the force chains.
Such studies on the drag force in granular media are relevant to understand the animal locomotion through granular media \cite{sharpe15, maladen09, aguilar16, hosoi15, maladen11, marvi14, pak11, mazouchova10, Li09}.
We expect that our study can stimulate the robotics or animal locomotion moving in sands.
  \begin{figure*}
  	\centering
  	\includegraphics[width=1.0\linewidth]{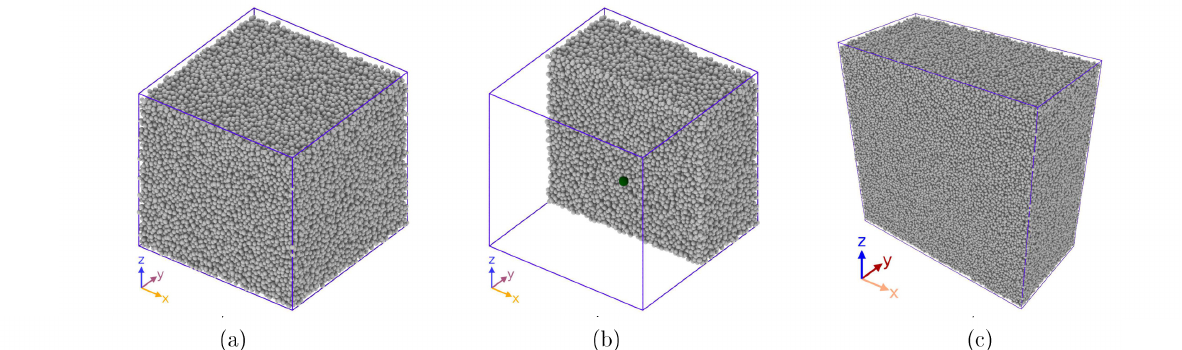}
  	\caption{The initial configurations for (a) system I with a size of $40d\times40d\times38d$: Initial configuration for the intruder diameter 2.0 $d$. Note that the intruder is located inside the system. (b) Slice through the plane $y=20.0d$ showing the intruder's position in the system for the case of $2.0d$ in system I. (c) System II of size $80d\times40d\times80d$ with the intruder diameter $4.0d$ positioned at a depth of $55d$ from the surface in the negative $z$ direction where the depth is varied for various cases while keeping the diameter of intruder constant.}
  	\label{fig:testing}
  \end{figure*}

Though many studies exists for characterization of the drag force, we need to establish a scaling law for the drag forces in granular media similar to the case of viscous fluid flows.
If there is an object in a viscous fluid and if there exists a relative speed $V$ between the object and the fluid, it is well-known that $F_{\rm drag}$ can be given as $(1/2)\rho C_\mathrm{D} A V^{2}$, where $A$ is a reference area, $\rho$ is the fluid density, and $C_\mathrm{D}$ is the drag coefficient.
Such kind of universal law does not exist for granular fluids.
In a recent study \cite{takehara14} for the drag force acting on a disk at high velocity near the jamming point, it is found that the average drag force can be expressed as $F_\textrm{drag}= F_{0}(\Phi) + \alpha(\Phi)V^{2}$, where $\Phi$ is the area fraction and $V$ is the pulling speed of the intruder disk.
This scaling  has also been observed in the numerical study \cite{takada17} of similar system.
Nevertheless, the situation is not simple for three-dimensional cases.
Indeed, the simulation in Ref.\ \cite{hilton} suggests $F_{\rm drag}$ for a pulling sphere in a three-dimensional granular media under the influence of the gravity is expressed as the summation of a yield force and the dynamical term which is proportional to $V$, at least, for $V<3\sqrt{g D/2}$, where $D$ and $g$ are, respectively, the diameter of the intruder and the  acceleration due to gravity.
On the other hand, an experiment on the moving cylinder in a three-dimensional granular media using Taylor-Couette configuration  \cite{reddy11} suggests that the drag law obeys $V\propto \exp[(F_{\rm drag}-F_{\rm c})/F_{\rm c} G]$, where $F_{\rm c}$ and $G$ are the yield force and a function that depends on the location (or the shear).
Therefore, we should clarify what is the proper scaling law to characterize the moving object in granular media in three-dimensional situations.

The recent work to study the effects of vibration on the drag force of an intruder revealed that the drag force does not depend on the velocity for low vibration level and depends linearly for high vibrations \cite{seguin17}.
Similarly velocity dependence of $F_{\rm drag}$ has been discussed in both experimental and numerical works \cite{hilton,albert99,geng05,baldassarri06,seguin16,potiguar13,faug15} for various shaped intruders and various packing conditions either in two-dimensional or three-dimensional systems.
Based on the results of these studies, it seems that there exist three or four regimes such as velocity independent (quasistatic), weakly dependent (logarithmic), linear, and quadratic regimes, though the distinction between quasistatic and the logarithmic regimes is not easy.
In the present work with the help of extensive computer simulations, we numerically analyze the drag force for a wide range of intruder velocities characterized by the Froude number defined by $\mathrm{Fr}={2 V}\sqrt{{2D}/{g}}\big/(D+d)$, where $d$ is the average diameter of the granular bed particles.
Plotting the numerical data against the Froude number, we clarify the existence of crossovers between these regimes.
Moreover we also attempt to know the role of depth of the immersed object on the crossover.
We discuss how the drag forces are scaled by the intruder diameter and the depth.
We also discuss how the scaling exponents depend on the friction coefficient.

The present study makes use of the simulation of pulling a spherical intruder in dense granular medium to answer the questions discussed above.
In Sec.\ \ref{sec:simu_method}, we explain the simulation methodology employed for the present study.
We illustrate that the drag force can be represented by a product of the yield force and the dynamical part.
We also discuss how the yield force depends on the intruder size, the depth of the intruder, and the friction coefficient in Sec.\ \ref{sec:resana}.
Finally we summarize our results with the possible future work in Secs.\ \ref{sec:discussions} and \ref{sec:CONCL}.

\section{Simulation methodology}\label{sec:simu_method}
We employ the discrete element method (DEM) as our simulation method \cite{cundall79}, which is widely used to simulate the granular materials.
The simulations are carried out for a three-dimensional system. Initial configurations for the simulation are generated by fixing the intruder sphere at a given position while granular particles of diameter uniformly distributed between $0.9d$ to $1.1d$ are poured from the top by the gravity.
The particles are allowed to settle and reach almost zero kinetic energy due to the presence of dissipating and damping forces.
In our simulations, physical quantities are scaled by the density $\rho$, the average diameter $d$, and the gravity acceleration $g$.
Therefore, for example, the time is scaled by $\sqrt{d/g}$, the mass is scaled by $\rho d^3$, the force is scaled by $\rho d^3 g$, etc.

Periodic boundary conditions are applied across $x$ and $y$-directions, where $x$ coordinate is assigned to the moving direction of the intruder, and $z$ is the vertical direction (anti-parallel to the direction of the gravity).
A rough base is used for the bottom of the simulation box.
The gravity acceleration acts along negative $z$-direction throughout the simulation.
We have considered two systems in our study.
The first system consists of $70,001$ particles (including one intruder) in a simulation box of $40d\times40d\times38d$ in the $x$, $y$, and $z$-directions.
The intruder is initially located at $x=y=20 d$ and $z=13.5 d$.
In this system, the depth of the intruder from the surface is kept constant while the diameter of the intruder is varied.
Six cases of the intruder diameter are considered with the intruder diameter varying from $1.0d$ to $6.0d$ at various moving velocities inside the granular bed.
It is ensured that the system is large enough for all the simulations carried out and that the periodic boundary does not affect the forces on the intruder due to its periodic image.
The initial configuration for the intruder of diameter $2.0d$ is shown in Figs.\ \ref{fig:testing}(a) and \ref{fig:testing}(b).
We have additionally considered the intruders of diameters $D=8d$ and $10d$ in the first system to confirm the scaling law on the yield force.
The second system consists of $300,001$ particles including the intruder in a system of $80d\times40d\times80d$ in the $x$, $y$ and $z$-directions (see Fig.\ \ref{fig:testing}(c)).
The intruder diameter is fixed as $4.0d$ while the depth of the intruder is varied as $10d$, $25d$, $45d$, $50d$, $55d$, $60d$, and $65d$ from the surface in the negative $z$-direction.
The first system I is used to study the dependence of drag force on the diameter of the intruder at constant depth for both frictional and frictionless cases, the drag force in the quasistatic regime and the yield force of the intruder (here, we define the yield force as the drag force acting on the intruder in the zero velocity limit).
The second system II, on the other hand, is used to study the drag forces on the intruder of a constant diameter $4.0d$ at various depths and the dependence of the drag forces on the other factors such as non-linear spring stiffness and damping coefficient both of which are defined in the next paragraph.
The benefit of using two systems is as follows.
The system I can be used to determine the yield force on the intruder because we need huge computation time (the number of particles in the system II is four times or more than that in the system I), while the system II is adequate to know the depth dependence of the drag force because the location of the intruder should be away from the bottom rough base and the surface of the granular bed.

The normal and tangential contact forces \cite{brilliantov96} between the two interacting particles or between the particle and the wall are expressed as:
\begin{align}
&F_n=\sqrt{R_{\textrm{eff}}\delta} (K_n\delta-m_{\textrm{eff}}\gamma_n v_n),\label{hertzianeq_n}\\
&F_t=-\textrm{min}(\mu F_n,\sqrt{R_{\textrm{eff}}\delta} (K_t\Delta {s_t}+m_{\textrm{eff}}\gamma_t v_t)),\label{hertzianeq_t}
\end{align}
Here, $K_n$ and $K_t$ are non-linear spring constants for the normal and the tangential contact forces, respectively, with the dimensions of the pressure, $\gamma_n$ and $\gamma_t$ are damping factors for the normal and the tangential forces, respectively, $\delta$ refers to the overlap distance of the colliding particles and is given by $\delta=(d_1+d_2)/2-|d_{12}|$ where $d_1$ and $d_2$ are the diameters of the colliding particles 1 and 2, respectively, and $d_{12}$ is the instantaneous distance between the two colliding particles.
Moreover, $\Delta s_t$ is the tangential displacement between two particles when in contact, $v_n$ and $v_t$ are the normal and the tangential components of the relative velocity, respectively.
$R_{\textrm{eff}}$ is expressed as $d_1d_2/(2(d_1+d_2))$ and the reduced mass $m_{\textrm{eff}}$ is expressed as $m_{\textrm{eff}}=m_1m_2/(m_1+m_2)$, where $m_1$ and $m_2$ are the masses of the colliding particles 1 and 2, respectively.
Here $\mu$ is the coefficient of friction, for particle-particle interactions, which is varied from $0.0$ to $0.9$ in the present simulation study.
There is an upper limit $\mu F_n$ on the tangential component of contact force to express the slip motion between two contacting particles.

\begin{figure}
    \centering
    \includegraphics[width=0.8\linewidth]{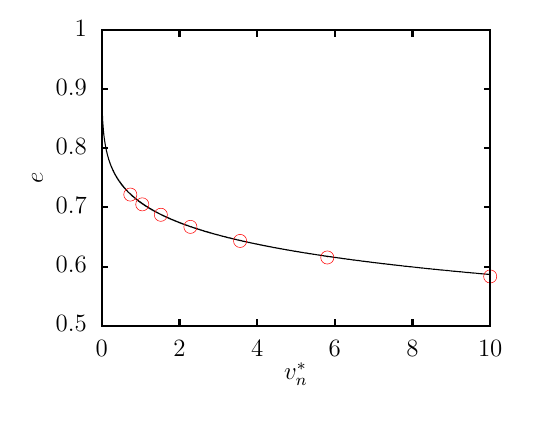}
    \caption{The coefficient of restitution $e$ for a granular material of diameter $d$ colliding with a flat wall of similar elastic properties.
    The open circles represent the coefficient of restitution obtained by our simulation while the curve of $e$ (solid line) represents the theoretical expression of $e$ by Kuwabara and Kono \cite{kuwabara87} using the first four constants from the work of Ramirez et al. \cite{ramirez99}.
    Here, we introduce the dimensionless velocity in the normal direction as $v_n^*\equiv v_n / \sqrt{gd}$.}
    \label{fig:restitu}
\end{figure}

We have taken $K_n=2\times10^8 \rho dg$ and the value of $K_t=2.45\times10^8 \rho dg$ for our simulations.
These values of $K_n$ and $K_t$ correspond to Young's modulus, $E=200\ {\rm GPa}$ and $\nu=0.3$ if we choose $\rho=10\ {\rm g/cm}^3$, $d=7.5\ {\rm mm}$, and $g=9.8\ {\rm m/s}^2$.
The identical $K_n$ and $K_t$ can be used to represent a material of Young's modulus, $E=10\ {\rm GPa}$ and $\nu=0.3$ if we choose $\rho=1\ {\rm g/cm}^3$, $d=3.75\ {\rm mm}$, and $g=9.8\ {\rm m/s}^2$.
The value of $\gamma_n$ is chosen as $50000 \sqrt{g/d^3}$ to obtain a coefficient of restitution curve that represents a realistic value.
It is seen that the coefficient of restitution depends on the colliding velocity \cite{kuwabara87}.
We have carried out the simulation of dropping a sphere on a flat wall to obtain coefficient of restitution for various $V_n$ as shown in Fig.\ \ref{fig:restitu} with the interaction parameters identical to that of the sphere-sphere interaction.
The theoretical value of the coefficient of restitution is also plotted (calculated taking diameter of wall as infinity) using the well-known expression obtained by Kuwabara and Kono \cite{kuwabara87} with the first four constants of the infinite series \cite{ramirez99,brilliantov96,schwager98} and is seen to converge with the one obtained from our simulations at various velocities (see Fig.\ \ref{fig:restitu}).
The coefficient of restitution in our simulation satisfies the well known curve of the form $1-k_1 (v_n)^{1/5}+k_2 (v_n)^{2/5}+...$, where $k_i$ are constants for a given $K_n$ and $\gamma_n$, and $v_n$ is the normal component of velocity along the line of impact.
Note that the coefficient of restitution curve remains the same if $\gamma_n/K_n^{0.6}$ ratio is kept constant \cite{schwager98}.
This has been used in some part of our study to maintain the same curve for the coefficient of restitution while varying the value of $K_n$.
Basically, we fix $K_n$ as $K_n=2\times10^8$ $\rho d g$ except for the places where we change $K_n$ explicitly.
The value of $\gamma_t$ is identical to $\gamma_n$ \cite{silbert01} in our simulation.
\begin{figure}
    \centering
    \includegraphics[width=.95\linewidth]{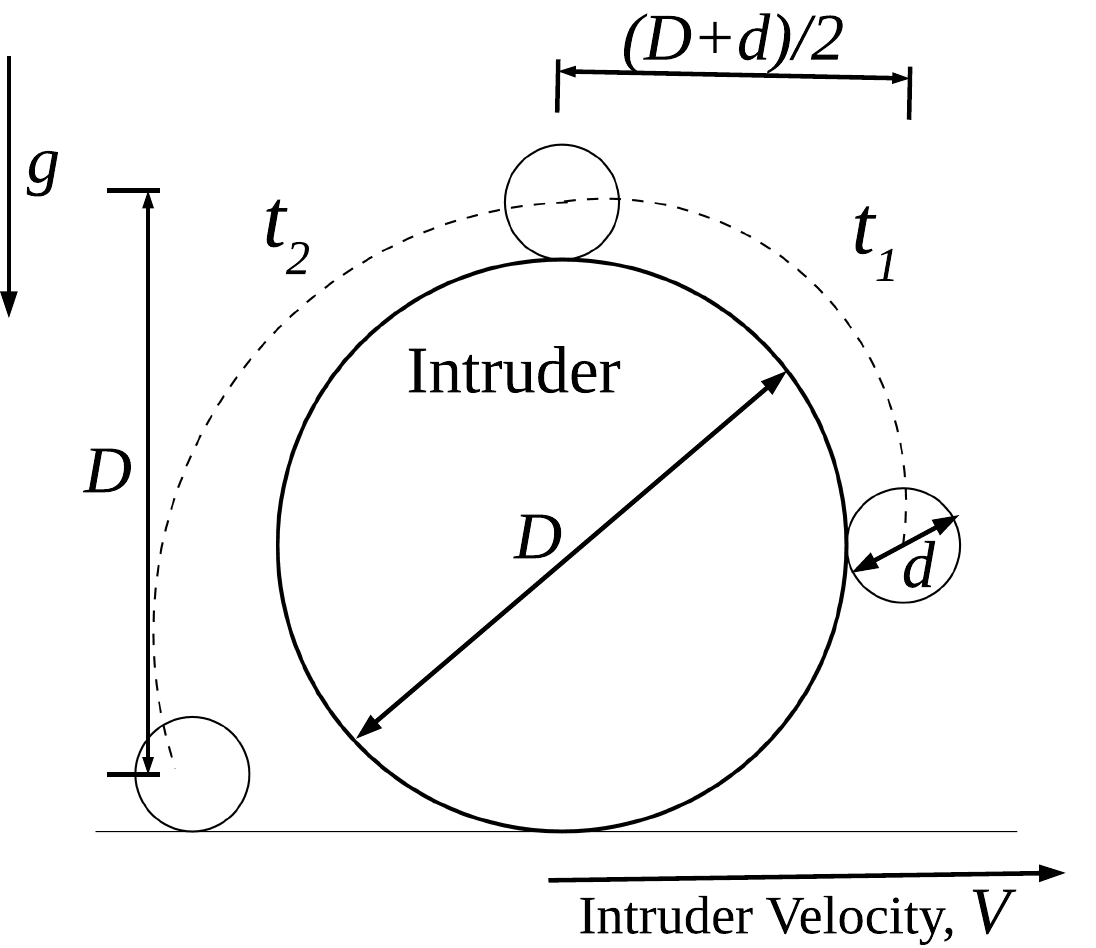}
    \caption{We define Froude number as the ratio of the two timescales, $t_2$ and $t_1$, as shown in the figure.}
    \label{fig:fr_fig}
\end{figure}

The value of a time step is chosen as $1.0\times 10^{-5} \sqrt{d/g}$ \cite{schwager98,schwager08}.
The intruder is moved with a constant speed in $x$-direction at a constant depth.
We have two time scales $t_1$ and $t_2$ in our system as shown in the Fig.\ \ref{fig:fr_fig}.
The first time scale $t_1$ characterizes the forward motion of the intruder and is given by $t_1=(D+d)/2V$.
The second time-scale $t_2$ represents the characteristic time for the particle to fall from the top to the bottom, which is given by $t_2=\sqrt{2D/g}$.
The Froude number $\mathrm{Fr}$ introduced in Sec.\ \ref{sec:intro} corresponds to:
\begin{equation}
\label{Fr}
\mathrm{Fr}=t_2/t_1=\frac{2V}{(D+d)}\sqrt{\frac{2D}{g}}.
\end{equation}

We are interested in this definition of Froude number because of the transition that happens at $t_1=t_2$, i.e., $\mathrm{Fr}=1$.
The simulations are performed for $\mathrm{Fr}\le 10$.
The coefficient of the friction $\mu$ is assumed to be a constant for all contacts.
The standard velocity Verlet time  integration algorithm  is used to update the particles positions and velocities in our DEM simulation.
The simulation is carried out until the intruder to cover the whole simulation box length, i.e., to move one complete length of the simulation box in the $x$ direction.
The simulations are carried out using Large-scale Atomic/Molecular Massively Parallel Simulator \cite{plimpton95,plimpton07}. OVITO \cite{stukowski10} and VMD \cite{humphrey96,kohlmeyer17} are used for post-simulation visualization and analysis.
\begin{figure}
	\centering
	\includegraphics[width=0.8\linewidth]{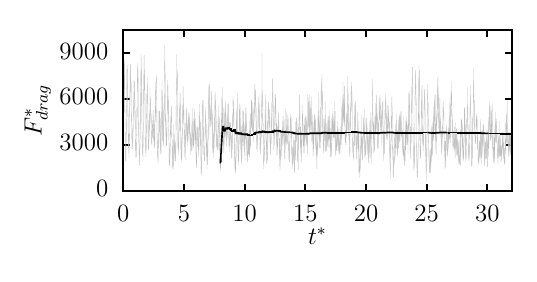}
	\caption{This figure shows instantaneous drag force on the intruder (thin line) and its cumulative average with respect to time (thick line), where we have introduced the dimensionless time $t^*\equiv t \sqrt{g/d}$, and the dimensionless drag force $F_{\rm drag}^*=F_{\rm drag}/(\rho g d^3)$.
		Here, the intruder diameter $D=4.0d$ and its moving speed is $1.0\sqrt{gd}$.
		This plot is based on the simulation of system I for $\mu=0.5$.}
	\label{fig:Flift_Fdrag_scheme}
\end{figure}

\section{\label{sec:resana} Results and Analysis}
\begin{figure*}
	\centering
	\includegraphics[width=0.95\linewidth]{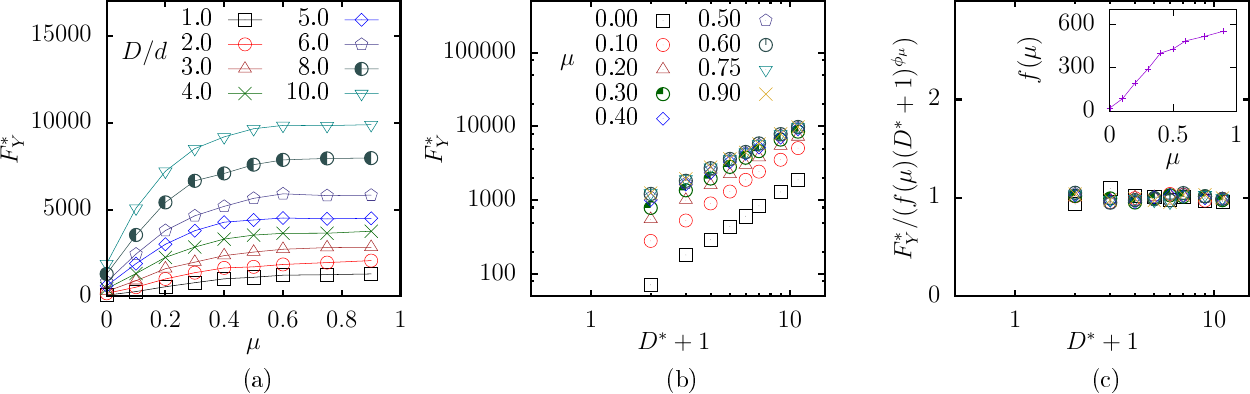}
	\caption{(a) The dimensionless yield force $F_{\rm Y}^*$ acting on intruders of various diameters $D$ with respect to coefficient of friction $\mu$. (b) The dimensionless yield force $F_{\rm Y}^*$ acting on intruders versus $D^*+1$ for the various values of coefficient of friction.
		We can see that the yield force satisfies $F_{\rm Y}^* \sim (D^*+1)^{\phi_\mu}$, where the exponent $\phi_\mu$ depends on $\mu$.
		(c) The scaled plot of the yield force, all the symbols have the same meaning as in (b). Inset of (c) shows friction coefficient dependence of $f(\mu)$.}
	\label{fig:yield_drag}
\end{figure*}
\begin{figure}
	\includegraphics[clip,trim=0cm 0.5cm 0cm 0.2cm,width=0.8\linewidth]{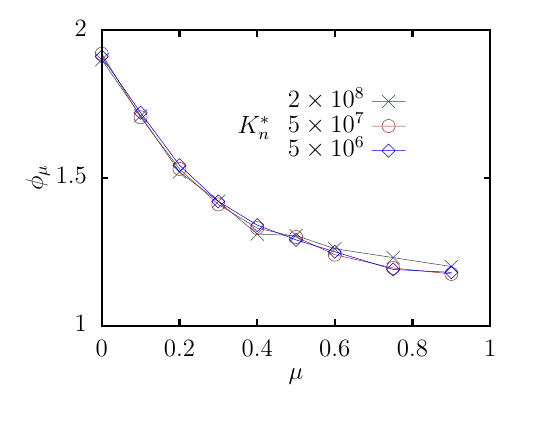}
	\caption{The variation of $\phi_\mu$ with respect to $\mu$ for various values of $K_n$.
		The coefficient of restitution curve and $K_t/K_n$ was kept constant while $K_n$ was varied.}
	\label{fig:phi}
\end{figure}

This section is the main part to present the results of our simulation study.
This section consists of 8 subsections.
In Sec.\ \ref{sec:yield}, we study the yield force on an intruder and find a scaling law for the yield force for several $\mu$ and $D$.
In Sec.\ \ref{sec:alphamuscale}, we study the yield force on an intruder for various $\mu$ at different intruder depth $h$ and propose an exponential scaling for the depth dependence.
In Sec.\ \ref{sec:fric}, we show the results of the drag force at $\mu=0.5$ for various intruder diameters $D$.
In Sec.\ \ref{sec:fricless}, we present the results of our simulation on the drag force for frictionless grains ($\mu=0$) for various intruder diameters.
In Sec.\ \ref{sec:quasis}, we examine how the drag force depends on the friction coefficient $\mu$ in the range $0\le \mu \le 0.3$ for $\mathrm{Fr}\le 1$.
In Secs.\ \ref{sec:drag-height}, we present the detailed results of the drag force for various depths for a constant intruder size $D=4.0d$.
In Sec.\ \ref{sec:kn}, we examine how the drag force depends on the stiffness constant $K_n$ and the dissipation constant $\gamma_n$.
In Sec.\ \ref{sec:scaling_D}, we demonstrate the existence of a scaling law for the drag force with respect to the yield force. We have plotted the instantaneous drag force acting on the intruder at a given time as shown in Fig.\ \ref{fig:Flift_Fdrag_scheme} for $D=4.0d$ and the pulling speed $1.0\sqrt{gd}$. Although the instantaneous force (thin line) acting on the intruder fluctuates with time, the cumulative average of this drag force is well defined (thick line).
Then, we call this average force  the drag force on the intruder in the subsequent subsections.

\subsection{Yield force and its dependence on $\mu$ and the intruder size $D$}\label{sec:yield}

In the present study, we have introduced the yield force as the drag force on an intruder moving through granular bed in the zero limit of $\mathrm{Fr}$.
As can be seen in the velocity field around the intruder (Fig.\ \ref{fig:vel-all}), there is a bulk reorganization of the particles around the intruder.
To obtain the yield force, we perform simulations at $V=0.01\sqrt{gd}$ and regard the average force acting on the intruder as the yield force.
Note that it is computationally very expensive to move an intruder below this velocity.
The simulations are carried out for nine cases for various coefficients of friction $\mu= 0$, $0.1$, $0.2$, $0.3$, $0.4$, $0.5$, $0.6$, $0.75$ and $0.9$ for the intruder diameters $D/d=1$, $2$, $3$, $4$, $5$, $6$, $8$ and $10$ in this study using System I.

The dimensionless yield forces $F_{\rm Y}^*=F_{\rm Y}/(\rho d^3 g)$ with respect to $\mu$ and $D^*+1$ with $D^*=D/d$ are plotted in Figs.\ \ref{fig:yield_drag}(a) and (b), respectively.
As can be seen in these figures, the yield force increases with $\mu$. At any moment, the force acting on the intruder can be split into two components: the normal forces due to the $K_n$ part and the tangential forces due to the $K_t$ part.
Here, the velocity dependent damping forces whose components are proportional to $\gamma_n$ and $\gamma_t$ does not play important role for small $\mathrm{Fr}$.
However, we notice that the contact force between the intruder and the particle in contact increases several times as $\mu$ increases from zero to a finite value.
It should be noticed that the yield force starts to saturate beyond a certain coefficient of friction $\mu$.
This saturation might be related to the upper limit of tangential force in the form of $\mu F_n$ where $F_n$ is the normal component of force, i.e., if the tangential force exceeds this particular value, then the tangential force becomes constant.
However, this switching of force does not take place for high $\mu$ because the tangential force does not become as large as $\mu F_n$. Hence, the yield force saturates after a certain value of $\mu$.

\begin{figure}
	\includegraphics[clip,trim=0cm 0.5cm 0cm 0.2cm,width=0.8\linewidth]{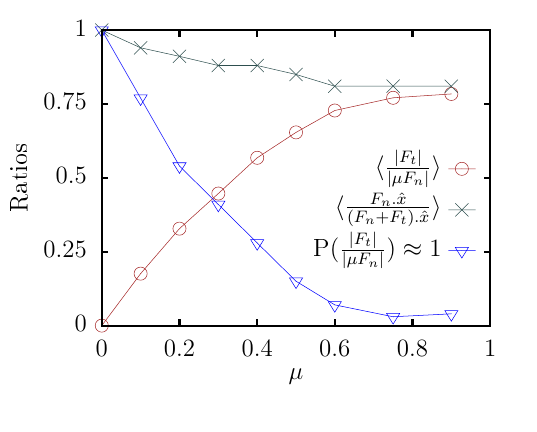}
	\caption{The variation of various ratios with respect to $\mu$. The first ratio $\big<\frac{|F_n|}{|\mu F_t|}\big>$ is the averaged ratio of tangential and $\mu$ times the normal force between the intruder and the particles in touch. Similarly, the second ratio $\big(\frac{F_n . \hat{x}}{(F_n + F_t).\hat{x}}\big)$ represents the contribution of normal collision force in the drag where $\hat{x}$ is the unit vector in $x-$direction. The third ratio , P$\big(\frac{|F_n|}{|\mu F_t|}\big)\approx1$ represents the fraction of particles in contact with intruder undergoing slipping. These plots are for $D=2.0d$.}
	\label{fig:ratio}
\end{figure}

To understand the mechanism behind the saturation of this yield force with $\mu$, we have considered three different parameters. First is $\big<\frac{|F_t|}{|\mu F_n|}\big>$ which represents the time-averaged ratio of tangential and $\mu$ times the normal collision force for each intruder contact. Second is $\big<\frac{F_n.\hat{x}}{(F_n+F_t).\hat{x}}\big>$ which is the fraction of drag force that comes from the normal collision force ($\hat{x}$ is the unit vector along the moving direction of the intruder). It can be seen in Fig. \ref{fig:ratio} that the second parameter decreases while the first one increases with $\mu$. Moreover, we have plotted the third parameter ($P(\frac{|F_t|}{|\mu F_n|}\approx1$) which represents the fraction of particles in contact with the intruder undergoing slipping. It can be seen that only around $ 5 \%$ of the particles undergo slipping at a high $\mu$. This is in accordance with the saturation of $F_Y$ with $\mu$ we discussed earlier. Importantly, we also see that the contribution of normal force to the drag changes from $100 \%$ at the frictionless case to approximately $80 \%$ at high $\mu$ although the increase in drag occurs several times than that. This proves that the presence of drag not only increases the frictional forces, but also the normal forces. This is possible as the particle in collision with the intruder shall be held strongly by the particles supporting it if $\mu$ is high which will also lead to a stronger force chain network.

\begin{figure}
	\centering
	\includegraphics[clip,trim=0.1cm 0.1cm 0.5cm 0.2cm,width=0.8\linewidth]{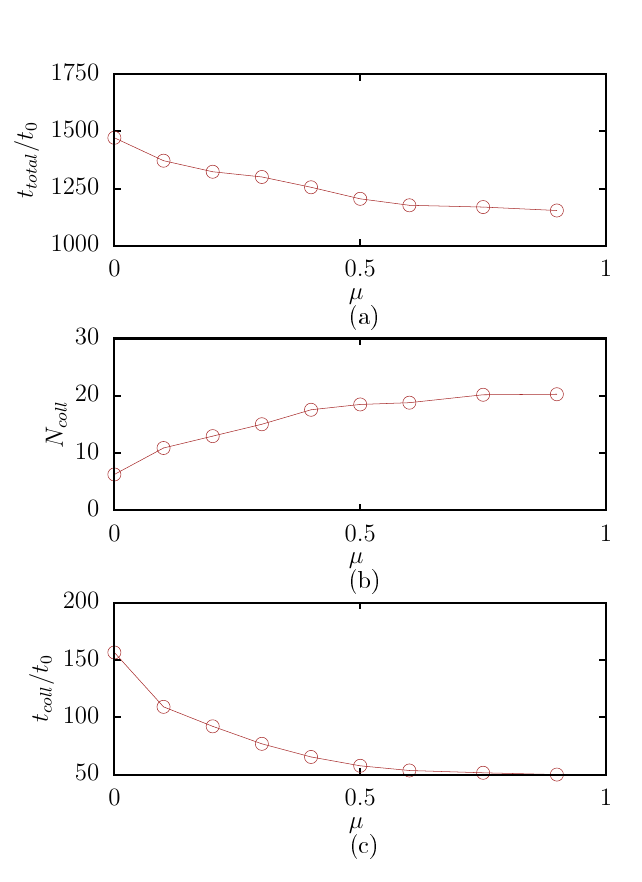}
	\caption{(a) Overall time a particle spends in contact with the intruder if it has made at least contact once, (b) the average number of times a particle collides with the intruder and (c) the average time per collision versus $\mu$. Here, $t_0$ is the collision time for elastic head-on collision two particles of diameter $d$ at the same velocity as that of intruder.\cite{schwager98}. These plots are for $D=2.0d$.}
	\label{fig:time_yield}
\end{figure}

\begin{figure*}
	\includegraphics[width=0.95\linewidth]{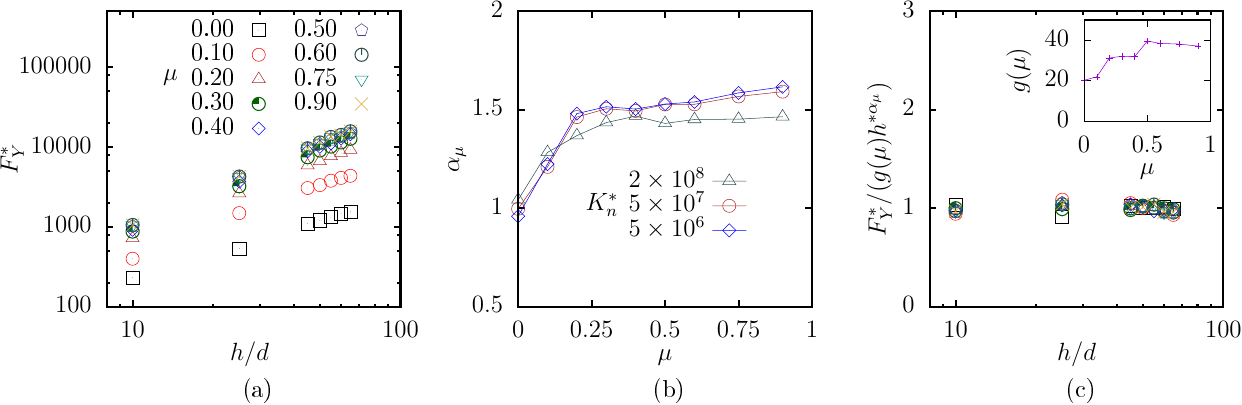}
	\caption{(a) The depth dependence of the yield force.
		Here, $h^*(\equiv h/d)$ is the dimensionless depth of the intruder from the surface in the negative $z$-direction.
		(b) Variation of $\alpha_\mu$ obtained from (a) with respect to $\mu$ for various non-linear spring constant $K_n$.
		(c) The scaled plot of the yield force, all the symbols have the same meaning as in (a) while the inset shows the friction coefficient dependence of $g(\mu)$.}
	\label{fig:alpha_gener}
\end{figure*}

Next, we plot the yield force against $D^*+1$ in a log-log plot in Fig.\ \ref{fig:yield_drag}(b).
The reason we have used $D^*+1$ instead of $D^*$ is that we want to plot the yield force against the effective diameter instead of just the intruder diameter.
In our work, the effective diameter is twice as much as the maximum distance between the intruder and a particle in the granular bed (of the average diameter $d$) that are in contact.
We find that yield force is proportional to $(D+d)^{\phi_\mu}$ where $\phi_\mu$ is a constant for a given coefficient of friction estimated from Fig.\ \ref{fig:yield_drag}(b), where the dimensionless yield force $F_{\rm Y}^*$ satisfies the scaling $F_{\rm Y}^*/(f(\mu)(D^*+1)^{\phi_\mu})={\rm const.}$
We also plot $\mu$-dependence to express the function $f(\mu)$ used in Fig.\ \ref{fig:yield_drag}(b) in the inset of Fig.\ \ref{fig:yield_drag}(c).
The exponent $\phi_\mu$ decreases with $\mu$ as shown in the Fig. \ref{fig:phi}.
The exponent $\phi_\mu$ for various $K_n$ is examined while keeping $K_t/K_n$ (which is a function of the Poisson's ratio only), $\gamma_n/K_n^{0.6}$ (the term that characterizes coefficient of restitution curve), and $\gamma_t/\gamma_n$ (the ratio of tangential and normal dissipation constants) as constants.
We can indicate that the exponent $\phi_\mu$ is almost independent of $K_n$ if we fix the other simulation parameters except $\mu$.

Note that $\phi_{\mu=0}\approx 2$ in the frictionless limit can be understood if the yield force is proportional to the cross section of the effective sphere surrounding the intruder.
In the absence of the frictional forces, the gravity plays the major role i.e., the intruder moves away from surrounding grains which are settled in a stable position by the gravity.
The work done by the intruder for changing the potential energy of the intruder is not retrieved back leading to a finite yield force for $\mu=0$.
In fact, if the intruder moves by a distance of $x$, all the particles inside a volume of $\pi (D+d)^2 x/4$ in front of the intruder has to be rearranged.
Since, in the zero velocity limit there is no momentum based collision forces or the frictional forces ($\mu=0$) thus leading to the square dependence for $\mu=0$.

It is important to understand how a given intruder contact displaces to make a way for the intruder. In a dense system with gravity such as ours, any given contact collides several times with the intruder. As the intruder moves, a given particle remains in contact for some time after which it separates. Due to gravity and a push from the particles at the back, the same particle may again be in contact with the intruder after some time. Let us say that $t_{total}$ is the total time a given particle spends in contact with the intruder summed for several small collisions, $t_{coll}$ is the average collision time for each contact (averaged for each such small collision) and $N_{coll}$ is the average number of collisions assuming the particle has collided at least once with the intruder. We see that although $t_{total}$ decreases with $\mu$, $N_{coll}$ is seen to increase with $\mu$ several times (refer to Fig. \ref{fig:time_yield}). Moreover, $t_{coll}$ is also seen to decrease with $\mu$. The decrease in $\phi_\mu$ can be a consequence of a change in these parameters.  In Fig. \ref{fig:time_yield}, it can be seen that although the total number of collisions increases with $\mu$, the overall time of contact $t_{coll}$ decreases, thereby implying that as $\mu$ increases, the intruder experiences an overall less time of contact with the other particles in the system. This leads to deviation from a perfect square dependence of the drag on the Froude number.

\subsection{Yield force dependence on $\mu$ and the intruder depth $h$}\label{sec:alphamuscale}
In this subsection, we have considered the depth dependence of the yield force.
As shown in Fig.\ \ref{fig:alpha_gener}(a), we have verified that the yield force is proportional to $h^{\alpha_\mu}$ for $h/d=10$, $25$, $45$, $50$, $55$, and $60$ using system II.
Indeed, $F_{\rm Y}^*/h^{*\alpha_\mu}$ is independent of $h/d$ as shown in Fig.\ \ref{fig:alpha_gener}(c), which can be scaled as a universal horizontal line with the introduction of $g(\mu)$.
We also plot $\mu$-dependence of the $g(\mu)$ in the inset of Fig.\ \ref{fig:alpha_gener}(c).
It is noted that the function $g(\mu)$ is different from $f(\mu)$ used in Sec.\ \ref{sec:yield} (see inset of Fig.\ \ref{fig:yield_drag}(c)).
Similar to the case with $\phi_\mu$ it is possible to scale the yield force with respect to the depth of the intruder. Then we study the scaling factor $(D+d)^{\phi_\mu}$ for various $h$ and $D$ for few cases to confirm that $\phi_\mu$ is independent of depth. With this in mind, we find that there exists a depth scaling factor $\alpha_\mu$ as ${F_\textrm{Y}} \propto h^{\alpha_\mu}{(D+d)^{\phi_\mu}}$.

\begin{figure*}
	\centering
	\includegraphics[width=0.95\linewidth]{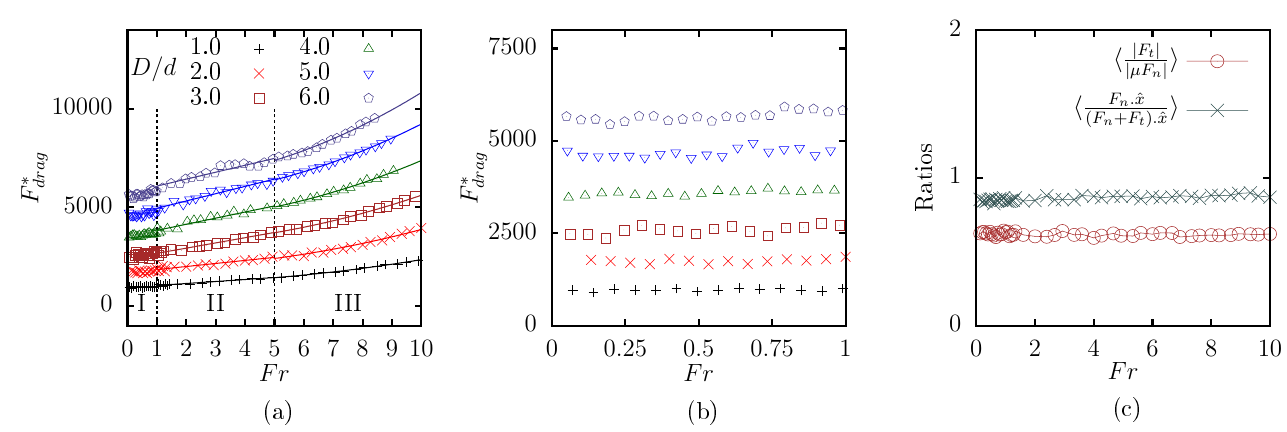}
	\caption{The drag force acting on the intruder of different diameters for various Froude numbers for $\mu=0.5$.
		A linear fit has been applied for $1<\mathrm{Fr}<5$ while a quadratic fit has been applied for $\mathrm{Fr}>5$.
		(a) Linear plot between drag force and $\mathrm{Fr}$.
		(b) Enlarged graph for quasistatic regime (Legend for bottom graph is same as top one). (c) Variation of the two parameters discussed in yield study with respect to $\mathrm{Fr}$.}
	\label{fig:Fr_Fx}
\end{figure*}
We also examine the $\alpha_\mu$ for various $\mu$ and the non-linear normal spring constant $K_n$ for a constant value of $K_t/K_n$, $\gamma_t/\gamma_n$, and $\gamma_n/K_n^{0.6}$ at seven different depths in our study ($h/d=10$, $25$, $45$, $50$, $ 55$, $60$, and $65$).
Unlike the case with the scaling factor $\phi_\mu$, $\alpha_\mu$ increases as $\mu$ increases.
However, $\alpha_\mu$ weakly depends on the non-linear spring stiffness $K_n$ as shown in Fig.\ \ref{fig:alpha_gener}(b).
The simplest contribution of the depth is the hydrostatic pressure expressed by $\rho g h$ which is roughly true for frictionless case, i.e., $\alpha_{\mu=0}\simeq 1$.
Changing the value of the non-linear spring constant $K_n$ or the Young's modulus $E$ causes the change of $\alpha_\mu$ because the overlap between two granular particles $\delta$ increases as the $K_n$ decreases. Hence, we propose
\begin{equation}
\frac{F_Y}{\rho (D+d)^{\phi_\mu} h^{\alpha_\mu} g}=\Psi
\end{equation}
where $\Psi$ is a function dependent of $\mu$. We shall call $\Psi$ as the yield parameter.
\subsection{Drag force in frictional system}\label{sec:fric}
In this subsection, we present the results of the drag force acting on a moving intruder moving through the granular bed at various velocities (in the positive $x$-direction) for $\mu=0.5$ and various $D$ with a constant depth of the intruder.
All the results in this subsection are based on the simulation of the system I.



The drag force with respect to the Froude number for $\mu=0.5$ has been plotted in the Fig.\ \ref{fig:Fr_Fx}(a), where $F_{\rm drag}^*=F_{\rm drag}/(\rho d^3 g)$.
We find three regimes, I, II and III in the drag force trend.
The drag force in the regime I ($\mathrm{Fr}<1$) is almost independent of $\mathrm{Fr}$ as can be seen in Fig.\ \ref{fig:Fr_Fx}(b).
The force in this regime strongly depends upon the coefficient of friction $\mu$ (as shown in Secs.\ \ref{sec:quasis} and \ref{sec:yield}) and $D$.
We call this regime the quasistatic regime as in Ref.\ \cite{chehata03,hilton,guillard14}, though we cannot exclude the weakly velocity dependent drag from our data.
In Ref.\ \cite{hilton}, a similar quasistatic regime was observed for $\mathrm{Fr}<1$ for various $\mu$.
We will study the dependence of drag force in quasistatic regime on the coefficient of friction later.
In a different experimental setup in Ref. \cite{chehata03}, a quasistatic regime was observed for $\mathrm{Fr}<1$ though their definition of $\mathrm{Fr}$ is different from ours.
We observe that the drag force in this regime is almost identical to the yield force which is the minimum force to make the intruder move inside the granular medium in the  limit of zero $\mathrm{Fr}$.
This yield force is synonymous to the yield stress in Bingham fluid which is the minimum stress to begin to flow under the shear.
The yield force is originated from the local jamming and buckling of force chain networks inside the granular bed.

In the regime II or the intermediate regime, i.e., $1<\mathrm{Fr}<5$, it seems that the drag force increases linearly with $\mathrm{Fr}$.
Least square linear fit has been applied in the regime II as shown in the graph of the form $a+b \mathrm{Fr}$, where $a$ and $b$ are fitting parameters.
Note that the slight deviation from linearity observed for the intruder of diameter $D=6.0d$ is mostly because the rough bottom wall is only a few particle diameters apart from the intruder.
The regime II which nicely follows linear trend as a function of $\mathrm{Fr}$ has been observed for all cases in the depth study (Sec.\ \ref{sec:drag-height}) where the  system size of our simulation is larger and the wall effect can be suppressed by keeping intruder far away from the bottom wall.
This regime was also observed in Ref.\ \cite{hilton}.
This regime II can be considered as a transition zone from the quasistatic regime (the regime I) to the regime III as mentioned in the next paragraph.

In the regime III, i.e., $\mathrm{Fr}>5$, a quadratic dependence of drag force on $\mathrm{Fr}$ can be observed due to the direct momentum transfer by collisions of grains to the intruder.
We call this regime  the inertial regime. A quadratic fit of the form $a^\prime+b^\prime \mathrm{Fr}^2$ (where $a^\prime$ and $b^\prime$ are fitting parameters for the quadratic expression) has been applied in this regime.
In a similar work for a two-dimensional system, where an intruder disk is dragged at a constant velocity under gravity \cite{potiguar13}, a quadratic dependence of drag force on velocity of intruder was observed.
There was absence of quasistatic and linear regime at least for the range of parameters ($d=0.32\ {\rm cm}$, velocity $= 10.3$--$309\ {\rm cm/sec}$, depth of immersion of intruder $=3.75$--$37.5\ {\rm cm}$) they investigated.
The reason they missed  the first two regimes could be due to the limitation of number of simulation runs at the low Froude number. We also plot the first two parameters, $\big<\frac{|F_t|}{|\mu F_n|}\big>$ and $\big<\frac{F_n.\hat{x}}{(F_n+F_t).\hat{x}}\big>$,  discussed in Sec. \ref{sec:yield} using intruder diameter $D=2.0 d$ in Fig. \ref{fig:Fr_Fx} (c) and we observed that the two parameters remain the same as that of yield limit in the range of $Fr$ we studied. This is important as we shall see in scaling of the drag force in the last subsection.
\begin{figure}
	\centering
	\includegraphics[width=0.75\linewidth]{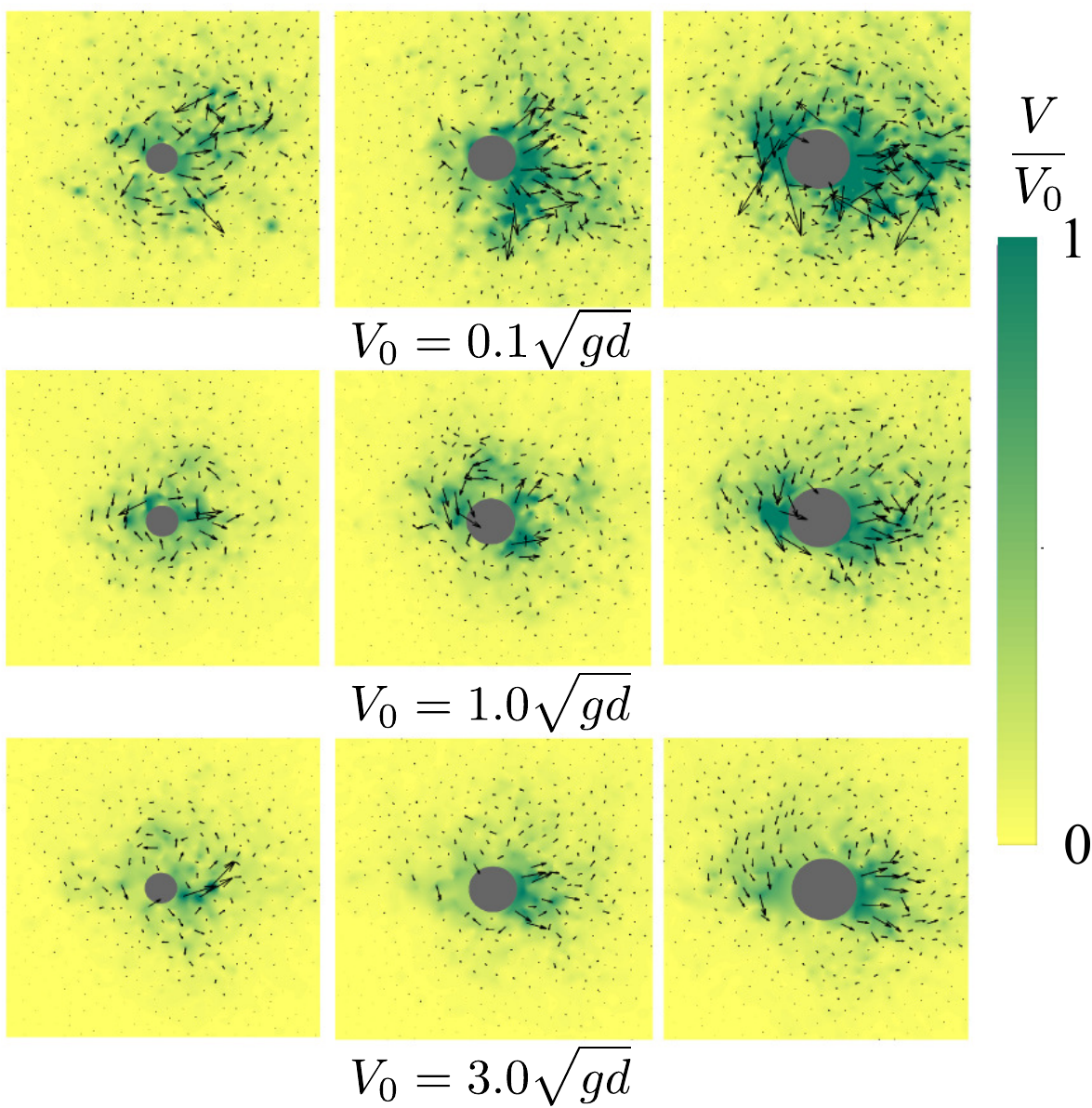}
	\caption{\label{fig:vel-all} Velocity field near the intruder at a particular instant of time for three intruders at three different velocities $V_0$ of $0.1$, $1.0$, and $3.0\sqrt{dg}$. The diameters of the intruder are $2.0d$, $3.0d$, and $4.0d$ respectively, in the order from left to right.}
\end{figure}

Figure \ref{fig:vel-all} shows the velocity field of the surrounding particles around the intruder for various $D$ and the moving speed $V_0$.
One can see that the surrounding particles for $V_0=0.1 \sqrt{gd}$ have more particles moving at speeds comparable to that of the intruder than for $V_0=1.0 \sqrt{gd}$.
The same trend can be observed between the intruder velocities $1.0\sqrt{gd}$ and $3.0\sqrt{gd}$.
Figure \ref{fig:vel-all} shows the larger number of rearrangements at lower velocity around the intruder than that for larger $V_0$.

\subsection{Drag force in frictionless system}\label{sec:fricless}
In the presence of the friction, the drag force is determined by two forces, the normal contact force $F_n$ and the tangential contact force $F_t$.
In the presence of the tangential contact forces, stick-slip motion in granular media plays an important role, i.e., when the intruder is in contact with a particle, the tangential contact force keeps building up (stick) until it reaches a value of $\mu F_n$, after which it starts slipping.
This allows the strain to be built up and then release in the form of rearrangements.
This phenomenon is quite active in the quasistatic regime. It is necessary to see how the drag force changes in the absence of these tangential contact forces.
\begin{figure}
	\centering
	\includegraphics[width=0.8\linewidth]{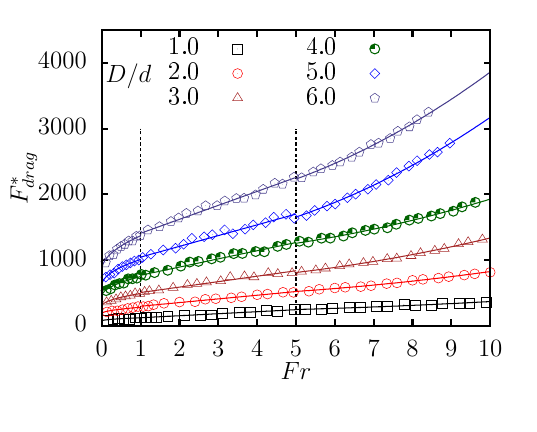}
	\caption{The average drag force acting on the intruder of different diameters for various Froude numbers for $\mu=0$.
	In the quasistatic regime, the data can be well fitted by $F_{\rm drag}^*=a^{\prime\prime}+b^{\prime\prime} \times {\rm Fr}^{0.5}$.}
	\label{fig:Fr_Fx_frictionless}
\end{figure}
\begin{figure}
	\centering
	\includegraphics[width=0.8\linewidth]{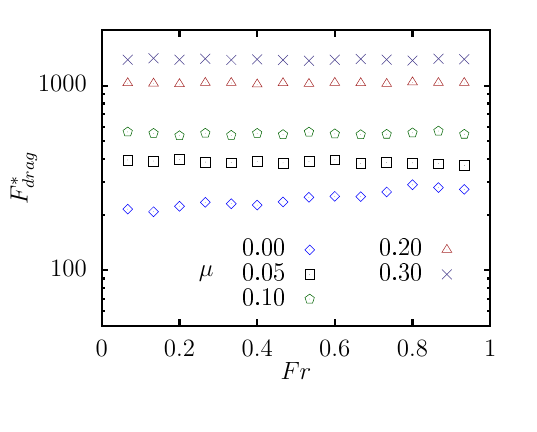}
	\caption{The average drag force acting on the intruder of diameter $2.0d$ in quasistatic regime for various $\mu$.}
	\label{fig:quasistatic_drag}
\end{figure}

All the simulations are carried out for the system I with $\mu=0$ for all the contacts.
We have obtained the result for the drag force on the intruder for frictionless cases as shown in Fig.\ \ref{fig:Fr_Fx_frictionless}.
Even in the frictionless case, three regimes exist.
However, contrary to the case with friction with a constant force, the drag force in the first regime can be fitted by the form $a^{\prime\prime}+b^{\prime\prime}\mathrm{Fr}^{0.5}$ as shown in Fig.\ \ref{fig:Fr_Fx_frictionless}.
The reason for the absence of the quasistatic regime here is that the velocity dependent  dissipative forces play dominant roles throughout the frictionless case because of the absence of the tangential contact force.
In other words, the main contribution is from the frictional force in this regime in the presence of friction.
However, we should stress that there exists a finite yield force even for the frictionless case.
This indicates that tangential force alone is not responsible for the yield force.
Nevertheless, we should note that the presence of tangential forces does increase the yield force multiple times (discussed earlier).
In the regimes II and III, we observe similar trends to those observed in the case with friction, i.e., we observe a linear and a quadratic dependences in II and III, respectively.
This is because the dominant force in these regimes is the normal contact forces as has already been discussed in the previous subsections.
The fact that the drag force follows the quadratic trend even in the case without friction also shows that the dominant force in this regime indeed is the collision based force and would be present even in the absence of the friction.
In the frictionless case, we observe that the crossover between the regimes (the drag force dependence on velocity of intruder) takes place at the identical Froude numbers as observed in the case with the friction.

\subsection{Drag force dependence on $\mu$ in the quasistatic regime}\label{sec:quasis}

In the previous subsections, we have reported the quasistatic regime only for the frictional case at $\mu=0.5$, while the quasistatic regime disappears in the frictionless case $\mu=0$.
Therefore, we should clarify whether a quasistatic regime exists for small $\mu$.
In this subsection, we study the drag forces in the quasistatic regime keeping the diameter of the intruder as $D=2.0 d$ for various $\mu$.

We have confirmed the existence of a constant force regime or a quasistatic regime for $\mathrm{Fr}\le 1$ except for $\mu=0$ as seen in Fig.\ \ref{fig:quasistatic_drag}, where we have examined the cases with $\mu=0$, $0.05$, $0.1$, $0.2$, and $0.3$.
Figure \ref{fig:quasistatic_drag} indicates the existence of the quasistatic regime even for the small coefficient of friction such as $\mu=0.05$.
We suggest that the drag force on an intruder in the quasistatic regime is constant unless $\mu$ is zero.
Another important quantity to be addressed is the yield force (the force required to initiate the motion of an intruder), which is  non-zero even for frictionless systems.
From Fig.\ \ref{fig:quasistatic_drag}, the drag force in this quasistatic regime is equal to yield force for the cases of $\mu \ge 0.05$.

\subsection{Drag force at various depth $h$}\label{sec:drag-height}
\begin{figure}
	\includegraphics[width=0.8\linewidth]{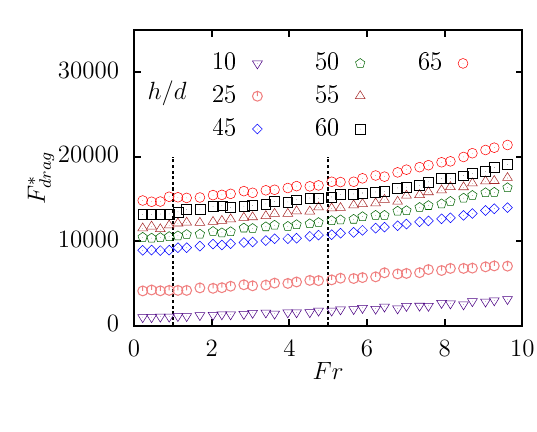}
	\caption{The dimensionless drag force $F_{\textrm{drag}}^*$ versus $\mathrm{Fr}$ for various depths. Here $h$ is the depth of the intruder from the surface in the negative $z$-direction.}
	\label{fig:height_drag}
\end{figure}
In a recent two-dimensional simulation study\cite{potiguar13} on a depth dependence of drag force on an intruder disk reports that the onset of inertial regime (the quadratic dependence of the drag force on $V$) clearly depends on the depth of the intruder.
Even in three-dimensional system, the onset of the quasistatic, the intermediate, and the quadratic regimes might depend on the amount of grains located above the intruder in granular media under gravity, which acts as a hydrostatic pressure proportional to $\rho g h$, where $h$ is the depth of the intruder.
Nevertheless, this simple picture might be only applicable to the frictionless case, because the exponent $\phi_\mu$ for $\mu>0$ is smaller than 2, and thus, we expect that the effective pressure is proportional to $h^{\alpha_\mu}$ with $\alpha_\mu>1$ for $\mu>0$.
Note that the sum rule $\phi_\mu+\alpha_\mu\approx3$ should be satisfied because of the argument on the yield force in the end of the last subsection.
Therefore, it is important to study the scaling of the drag force on the depth of the intruder.

All the results presented in this subsection are based on the simulations for the system II.
We have performed the simulations at depths $h/d=10$, $25$, $45$, $50$, $55$, $60$, and $65$ from the free surface in negative $z$-direction with the intruder diameter $D=4.0 d$ in a much larger system of dimensions $80d\times 40d\times 80d$.
The value of $\mu$ is fixed to be $0.5$, and all the other parameters such as $K_n, K_t, \gamma_n, \gamma_t$ are the same as mentioned in the simulation methodology section for the most of arguments in this subsection.
Based on Fig.\ \ref{fig:height_drag}, we have confirmed that the onset of the different drag regimes starts at roughly the identical Froude numbers as before.
It is reasonable that if there are insufficient numbers of granular layers above the intruder, the drag may undergo a regime change differently and the quadratic regime may start quicker as has been seen in Ref.\ \cite{potiguar13} for a two-dimensional system.
However, if the intruder is located sufficiently deep from the free surface, Froude number at which crossover between the regimes happens, seems to be almost independent of the depth.
Therefore, we expect the effect of the depth appears as the yield force.




\subsection{Drag force dependence on non-linear spring stiffness $K_n$, the dissipation coefficient $\gamma_n$, and the coefficient of restitution curve}\label{sec:kn}
\begin{figure}
	\includegraphics[width=0.8\linewidth]{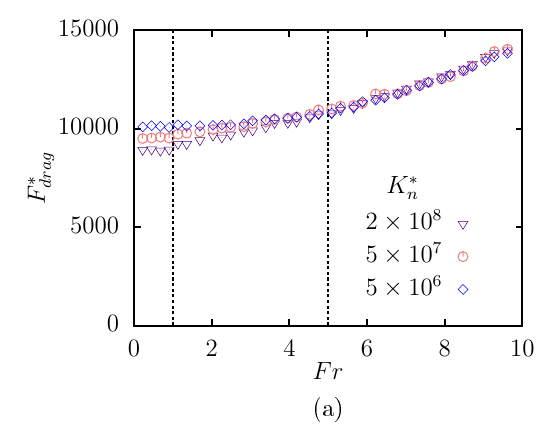}
	\includegraphics[width=0.8\linewidth]{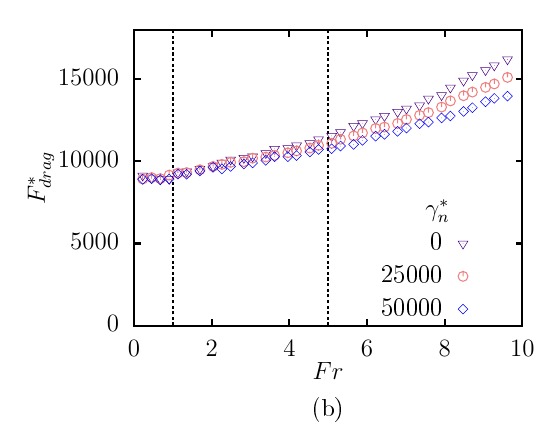}
	\caption{The variation of drag force with respect to $\mathrm{Fr}$ at fixed height for (a) various values of  $K_n$ and same coefficient of restitution curve and (b) various $\gamma_n$, different coefficient of restitution curve and constant $K_n$.}
	\label{fig:kn_kt_drag}
\end{figure}
\begin{figure*}
	\includegraphics[width=0.95\linewidth]{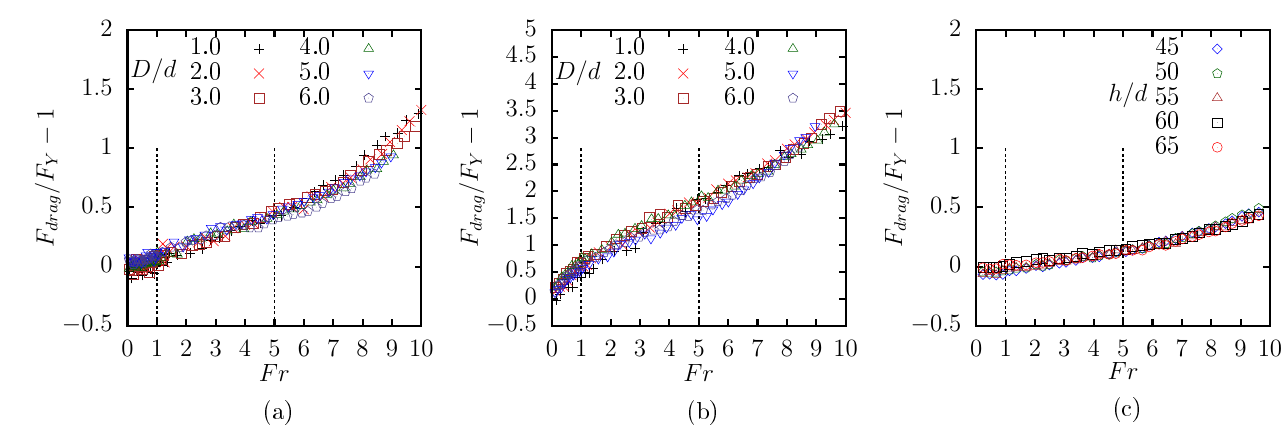}
	\caption{The scaled drag force against $\mathrm{Fr}$ for various intruder diameter $D$ for (a) $\mu=0.5$ and (b) $0$. (c) The scaled drag force against $\mathrm{Fr}$ for various intruder depths $h$ for $\mu=0.5$.}
	\label{fig:Fr_Fx_universal}
\end{figure*}

In this subsection, we study how the values of the non-linear spring stiffness constant $K_n$ changes the onset of regimes.
The simulations are performed at constant depth using the system II with $h=45d$.
We control $K_n$ with keeping $K_n/K_t$, $\gamma_n/K_n^{0.6}$, and $\gamma_n/\gamma_t$ constant.
Because $K_n/K_t$ is purely a function of the Poisson's ratio, we fix the Poisson's ratio for the material.
Similarly fixing $\gamma_n/K_n^{0.6}$ corresponds to the fixed curve for the coefficient of restitution $e$ (Fig.\ \ref{fig:restitu}) for the three cases of $K_n$ as shown in Fig.\ \ref{fig:kn_kt_drag}(a).
Finally, $\gamma_n/\gamma_t$ is kept equal to unity to maintain uniformity in the normal and tangential dissipation factor in all the three cases.

We observe that, all the graphs for $F_\textrm{drag}$ versus $\mathrm{Fr}$ seem to converge in the inertial regime, i.e., for $\mathrm{Fr}>5$ if we choose an identical restitution coefficient curve (by fixing the value of $\gamma_n/K^{0.6}_n$).
We, however, notice that the drag force depends on the spring stiffness in the first two regimes.
The drag force becomes smaller for larger $K_n$ in these regimes.
This is reasonable because the softer materials may form clusters (force chains) under the compression.
We know that the coefficient of restitution depends on the Young's Modulus $E$ or the normal spring stiffness $K_n$ \cite{kuwabara87}, but it is possible to vary the dissipation term $\gamma_n$ under unchanged curve of the coefficient of restitution.
This is what we have carried out by keeping $\gamma_n/K_n^{0.6}$ constant for various $K_n$ in Fig.\ \ref{fig:kn_kt_drag}(a) while we have varied $\gamma_n/K_n^{0.6}$ term in Fig.\ \ref{fig:kn_kt_drag}(b) for a fixed $K_n=2\times10^8 \rho d g$ and thus changing the coefficient of restitution curve.
As can be seen in our results, the drag force in the quadratic regime for various $K_n$ is similar as long as the $\gamma_n/K^{0.6}_n$ is kept constant irrespective of the value of $K_n$ (see Fig.\ \ref{fig:kn_kt_drag}(a)) and it is different for various $\gamma_n/K^{0.6}_n$ (Fig.\ \ref{fig:kn_kt_drag}(b)).
Hence, we can say that the forces in the quadratic regime depend on $\gamma_n/K^{0.6}_n$ but not on the normal non-linear spring constant $K_n$ or the normal dissipation $\gamma_n$ term separately.
We also observe that the forces in the quasistatic regime depend on $K_n$ but not on $\gamma_n$ as shown in Figs. \ref{fig:kn_kt_drag}(a) and \ref{fig:kn_kt_drag}(b).
This is to be expected as the damping terms containing $\gamma_n$ are associated with the velocity part and hence the drag independence on $\gamma_n$ in the quasistatic regime. This was also discussed in the yield force study (which lies in the quasistatic regime).


From both Figs.\ \ref{fig:kn_kt_drag}(a) and \ref{fig:kn_kt_drag}(b), we also conclude that the crossover in the drag regime that we have observed throughout our study is independent of the material properties such as $K_n$ and $\gamma_n$.
This is important as it sets a general rule for the regime changes in the drag force with respect to the Froude number used in our study, i.e., the drag force regime crossovers occur at $\mathrm{Fr}\approx 1$ between the quasistatic to the linear regimes and at $\mathrm{Fr}\approx 5 $ between the linear to the quadratic regimes.

\subsection{Scaling of the drag force with respect to the yield force}\label{sec:scaling_D}

In this subsection, we try to scale the drag force with respect to the intruder diameter and the depth. In the Sec. \ref{sec:fric}, we saw that the contribution of the normal collision force and the ratio of tangential and normal collision force remains constant against velocity of the intruder. Hence, indicating the presence of yield force in the scaling of the drag. Note that the data for the dynamical simulation with finite $\mathrm{Fr}$ are only for $D^*\le 6$.
In Sec.\ \ref{sec:yield}, we have found that the yield force can be scaled as $\rho (D+d)^{\phi_\mu} h^{\alpha_\mu}$, where $\phi_\mu$ and $\alpha_\mu$ obey Fig.\ \ref{fig:phi} and Fig. \ref{fig:alpha_gener}(b), respectively.
We see that the yield force can be used as a scaling for the drag forces up to $\mathrm{Fr}=10$, i.e., the scaled drag force $({F_\textrm{drag}}/{F_Y}-1)$ becomes independent of the size of the intruder (see Figs.\ \ref{fig:Fr_Fx_universal}(a) and \ref{fig:Fr_Fx_universal}(b)) as well as the depth (Fig. \ref{fig:Fr_Fx_universal}(c)).
If we are interested in the case for $D\gg d$, $F_\textrm{drag}\propto D^{\phi_\mu}$ for any given $\mathrm{Fr}$.
The existence of the scaling via the the yield force suggests that the drag force can be expressed as a product of the yield force and the dynamical part depending on $\mathrm{Fr}$.
Indeed, three regimes and their crossover values do not depend on $D$ and $\mu$.
We, therefore, propose that the drag force can be expressed as (for $\mu \ge 0.05$):
\begin{equation}
\label{eq:dragscale-friction}
\frac{F_{\textrm{drag}}}{F_Y}-1=
\begin{cases}
0, & \text{}\ \mathrm{Fr}<1 \\
\textrm{Linear} , & \text{}\ 1<\mathrm{Fr}<5 \\
\textrm{Quadratic} & \text{}\ \mathrm{Fr}>5, \\
\end{cases}
\end{equation}

We verify this by fitting a function of type $F_{drag}/F_Y-1=a_i\times \mathrm{Fr}^{b_i}$ on a log-log plot for the frictional cases in all the regimes where $a_i$ and $b_i$ are fitting parameters for the regime $i$. We see that $b_{III}$ is 1.7 and 1.9 for Fig. \ref{fig:Fr_Fx_universal}(a) and (c), respectively. In frictionless case, although the scalings are valid with respect to $F_Y$, these are applicable in regime II and III only when the fitting function is modified to the form $a_i\times \mathrm{Fr}^{b_i}+c_i$ where $c_i$ is additional fitting parameter.


\section{\label{sec:discussions} DISCUSSION}
The scaling law that we address in our work can play an important role in miniaturization of the drag forces on intruders to move through granular media.
The miniaturization of the drag in the granular media keeping the dimensionless number, would be important in industrial applications.
We, however, want to scale the diameter of the intruder $D$ by the diameter of the grains $d$, because the numerical simulation can treat only dimensionless quantities such as $D/d$.
This can be successfully used to do a complete miniaturization of the intruders on a laboratory scale while keeping the diameter of the granular bed particles same.
In the present work, we have found that the drag force can be represented by a product of the dynamical part and the yield force, if the intruder is located enough deep.
Thus, the scaling law for the moving intruder in frictional grains can be expressed by
\begin{equation}
\frac{F_\textrm{drag}}{\rho h^{\alpha_\mu}(D+d)^{\phi_\mu}g \Psi}=
\begin{cases}
1, & \text{}\ \mathrm{Fr}<1 \\
\frac{25}{4}\frac{\lambda}{\Psi} (\mathrm{Fr}-1) + 1, & \text{}\ 1<\mathrm{Fr}<5 \\
\frac{\lambda}{\Psi} \mathrm{Fr}^2 + 1, & \text{}\ \mathrm{Fr}>5 \\
\end{cases}
\label{eq:Fdrag_scaling}
\end{equation}
where the exponents $\alpha_\mu$ and $\phi_\mu$ satisfy the approximate sum rule $\alpha_\mu+\phi_\mu\approx 3$.
Unfortunately, we cannot explain $\mu$-dependences of the nontrivial exponents $\alpha_\mu$ and $\phi_\mu$.
The explanation of the exponents would be an important subject of our future work.

To express Eq.\ (\ref{eq:Fdrag_scaling}), we make the following assumptions:
(i) in the quasistatic regime, i.e., for $\mathrm{Fr}<1$, the force remains constant throughout and is equal to its yield force.
We have seen that this is not true for the case without friction.
However, for $\mu\ge0.05$, this is found to be true.
(ii) The graphs are continuous at both $\mathrm{Fr}=1$ and $\mathrm{Fr}=5$. This is what we have used to obtain the function in regime II using parameters from regime I and III ($\Psi$ and $\lambda$, respectively).
(iii) The intruder is not located in shallow region.
In other word, if the location of the intruder is shallow, we cannot use Eq.\ (\ref{eq:Fdrag_scaling}) because the drag force cannot be expressed as a product of the yield force and the dynamical part.

Moreover, based on our work in Sec. \ref{sec:yield}, \ref{sec:alphamuscale} and \ref{sec:kn}, we can say that \begin{equation}
\Psi\equiv \Psi(K_n^*,\mu)
\end{equation}
\begin{equation}
\lambda\equiv \lambda\bigg(\frac{\gamma_n^*}{K_n^{*{0.6}}}, \mu\bigg)
\end{equation}

We already saw in Fig. \ref{fig:yield_drag}(a) that $\Psi$ is heavily dependent on $\mu$, while Fig. \ref{fig:kn_kt_drag}(a) and (b) showed that it is weakly dependent on $K_n$ while is independent of $\gamma_n$. The weak dependence on $K_n$ was seen because the yield force could be affected by the ability of the particles to form force chains which is affected by the softness or hardness of the particles. The independence from $\gamma_n$ is a consequence of the fact that the damping term is coupled with the velocity component and thus gets nullified in the zero velocity limit. $\lambda$ is seen to be heavily dependent on all the parameters. However, since it is associated with the inertial regime, we can couple two of the parameters $\gamma_n^*/K_n^{*0.6}$ together since the forces remain same in inertial regime if this ratio is kept constant irrespective of the $K_n$ or $\gamma_n$ (see Fig. \ref{fig:kn_kt_drag} (a) and (b)).

\section{Conclusion}\label{sec:CONCL}
We have computed the drag force acting on an intruder moving with a constant velocity through the granular bed made up of slightly polydispersed particles. We have found that the drag force in the zero velocity limit could be scaled by the size and depth of the particle in a manner similar to that in fluids. We have seen how the parameters associated with these scalings change with frictional coefficient and other material properties. We have also investigated the saturation of yield force with frictional coefficient and related it to very low fraction of particles undergoing slipping at high $\mu$. Alternatively, we have also studied the time spent by an average particle as an intruder contact and its successive collisions with the intruder for various $\mu$. Also, we have related the depth scaling factor $\alpha_\mu$ with $\mu$ for various material properties. It was observed that $\phi_\mu+\alpha_\mu \approx 3$. Moreover, $\phi_\mu=2$ and $\alpha_\mu=1$ for frictionless system suggesting a projected area dependence like in fluids.

We have found the drag force can be expressed as a product of the yield force and the dynamical part, if the location of the intruder is sufficiently deep.
In the dynamical part of the drag force, there are three regimes: the regime I, the intermediate and, the inertial regimes, at least, for the frictional cases, where the crossovers take place at $\mathrm{Fr}\approx1$ and $\mathrm{Fr}\approx5$.
For $\mu\ge0.05$, the regime I for $\mathrm{Fr}\le 1$ is a quasistatic regime, while the drag force in the regime I for the frictionless case depends on $\mathrm{Fr}$.
Moreover, we have seen that the drag force in the inertial regime remains unchanged if the term that characterizes the coefficient of restitution curve, i.e., $\gamma_n/K^{0.6}_n$  is kept constant. A drag force in the velocity range we have studied can always be scaled with respect to the yield force.

\section*{ACKNOWLEDGMENT}
We would like to thank IITG for providing excellent computational facilities through Param-Ishan.
K.\ Anki Reddy would like to thank YITP, Kyoto University for funding one month visit.
This work is partially supported by the Grant-in-Aid of MEXT for Scientific Research (Grant No.\ 16H04025).

\bibliographystyle{plain}

\end{document}